\documentclass[aps,pre,showpacs,twocolumn,showemail]{revtex4}
\usepackage{bbm}
\usepackage{mathrsfs}
\usepackage{epsfig,psfrag}
\usepackage{amsmath,amsfonts,amssymb}
\usepackage[usenames]{color}
\usepackage[dvips, bookmarks, colorlinks=true, plainpages = false, citecolor = blue, linkcolor = blue, urlcolor = blue, filecolor = blue]{hyperref}
\def\be{\begin{equation}}
\def\ee{\end{equation}}
\def\bea{\begin{eqnarray}}
\def\eea{\end{eqnarray}}

\begin{document}

\preprint{draft}
\title{Scaling and Multiscaling Behavior of the Perimeter of
Diffusion$-$Limited Aggregation (DLA) Generated by the
Hastings$-$Levitov Method}

\author{F. Mohammadi $^1$}\email{mohammadi@physics.sharif.edu} \author{A. A. Saberi $^2$}\email{a$_$saberi@ipm.ir} \author{S. Rouhani $^1$ }

\address {$^1$ Department of Physics, Sharif University of Technology, P.O. Box 11155-9161,
Tehran, Iran. \\$^2$ School of Physics, Institute for Research in
Fundamental Sciences (IPM), P.O.Box 19395-5531, Tehran, Iran. }
\date{\today}

\pacs{64.60.al, 05.20.-y, 61.43.Hv, 68.35.Fx}

\begin{abstract}
In this paper, we analyze the scaling behavior of \emph{Diffusion
Limited Aggregation} (DLA) simulated by Hastings-Levitov method. We
obtain the fractal dimension of the clusters by direct analysis of
the geometrical patterns in a good agreement with one obtained from
analytical approach. We compute the two-point density correlation
function and we show that in the large-size limit, it agrees with
the obtained fractal dimension. These support the statistical
agreement between the patterns and DLA clusters. We also investigate
the scaling properties of various length scales and their
fluctuations, related to the boundary of cluster. We find that all
of the length scales do not have a simple scaling with same
correction to scaling exponent. The fractal dimension of the
perimeter is obtained equal to that of the cluster. The growth
exponent is computed from the evolution of the interface width equal
to $\beta=0.557(2)$. We also show that the perimeter of DLA cluster
has an asymptotic multiscaling behavior.
\end{abstract}

\maketitle
\section{introduction}

Diffusion-limited aggregation (DLA), introduced by Witten and Sander
\cite{Ref1}, has been shown to describe many pattern forming
processes including dielectric breakdown \cite{Ref2},
electrochemical deposition \cite{Ref3,Ref4}, viscous fingering and
Laplacian flow \cite{Ref5} \emph{etc}.\\ This model begins with
fixing a \emph{seed} particle at the center of coordinates in $d$
dimensions. By releasing random walkers from infinity and allowing
them to stick as soon as they touch the cluster, a fractal pattern
grows.\\ This procedure is equivalent to solving Laplace's equation
outside the aggregated cluster with appropriate boundary conditions.
The walker sticks to a point on the surface of the aggregate with a
probability proportional to the local field strength at that point
(the harmonic measure).

In two dimensions, since analytic functions automatically obey
Laplace's equation, the theory of conformal mappings provides
another mechanism for producing the shapes. This method has been
directly used by Hastings and Levitov (HL) to study DLA \cite{Ref6}.
These authors showed that DLA in two dimensions can be grown by
using successive iterating stochastic conformal maps. In the present
paper, we are interested in these \emph{off-lattice} DLA patterns
generated by this method.

We present some evidence that the patterns generated by HL method
have the same statistics as DLA clusters simulated according to the
original definition. In the first part of the paper, we calculate
the fractal dimension of the cluster patterns by direct
measurements. We use two different methods, first, using the scaling
relation between the average gyration radius of the generated
patterns with their size, and the second, calculating the density
two-point correlation function. We show that the results agree with
the fractal dimension of DLA clusters.

In the second part of the paper, we investigate the scaling
properties of various length scales and their fluctuations, related
to the boundary of the patterns. We examine whether they follow a
simple scaling relation with a same correction to scaling exponent,
or their scaling behavior is governed by the multiscaling property.

The multiscaling of DLA clusters, proposed by Coniglio and Zannetti
\cite{Ref7}, stands for space dependent fractal dimension according
which a whole set of scaling exponents exists. It has been also
claimed by Somfai, \emph{et.al.} \cite{Ref8,Ref9,Ref13}, that these
scaling claims are misled by finite size transients, and DLA obeys
simple scaling and all length scales scale with the same fractal
dimension.

However our simulation for clusters generated by HL method, shows
that the growth exponent defined by the interface width, differs
from the fractal dimension, and we find no correction to scaling
exponent for it. Furthermore, we extend the concept of multiscaling
to the boundary of the clusters and we find that the asymptotic
behavior of the boundary also agrees with the multiscaling property.

\section{The Hastings$-$Levitov Method\label{section HL}}

In the quasi-stationary approximation, the probability density of
finding a particle satisfies the Laplace equation
\begin{equation}\label{Eq1}
\nabla^2\psi(z)=0,
\end{equation}
with boundary conditions

\begin{equation}\label{Eq2}
                \psi(z)= \left\{ \begin{array}{cl}
                0 &  z\in\partial \mathcal{C}\\
                \frac{1}{2\pi}\ln|z| & |z|\rightarrow \infty,
              \end{array}\right.
\end{equation}
where the zero boundary condition on the boundary of cluster
$\partial \mathcal{C}$, implies the sticking of the particle upon
arrival, and the later condition states that $\psi(z)$ is
independent of any direction at infinity. \\The probability of
cluster growth at a certain point $z$ of the boundary of the cluster
is determined by the harmonic measure
\begin{equation}\label{Eq3}
    dP(z)=|\nabla \psi(z)| dl,
\end{equation}
where $dl$ is a boundary element containing the point $z$.

According to the Riemann mapping theorem, there exists a conformal
map that maps the exterior of the unit circle to the exterior of the
cluster. Hastings and Levitov constructed this map using the
iteration of conformal mapping \cite{Ref6}. The function
$\phi_{\lambda , \theta}(w)$ maps the unit circle to a circle with a
bump of linear size $\sqrt{\lambda}$ at the point $w = e^{i\theta}$,
\begin{equation*}
    \phi_{\lambda , 0}(w) =     w^{1-a}    \left\{\frac{1+\lambda}{2w}(1+w) \right.
\end{equation*}
\vspace{-20 pt}
\begin{equation}\label{Eq4}
    \left.   \left[1+w+w\left(1+\frac{1}{w^2}-\frac{2}{w}\frac{1-\lambda}{1+\lambda}    \right)^\frac{1}{2} \right]
  -1\right\}^a,
\end{equation}
\begin{equation}\label{Eq5}
    \phi_{\lambda , \theta}(w)=e^{i\theta}\phi_{\lambda
    ,0}\left(e^{-i\theta}w\right).
\end{equation}
The parameter $0\leq a\leq1$ determines the shape of the bump, for
higher $a$ the bump becomes elongated in the normal direction to
$\partial \mathcal{C}$, e.g. it is a line segment for $a=1$. In this
paper we set $a=\frac{1}{2}$ for which the bump has a semi-circle
shape.

\begin{figure}[t]
  \includegraphics[width=0.4\textwidth]{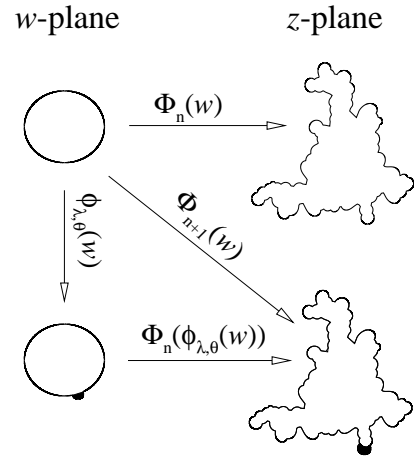}
  \caption{A circle in the $w$-plane is mapped to a
 $\mathcal{C}_n$ in the $z$-plane by $\Phi_n(w)$.
  The same function maps a circle with a bump at
  $\theta_{n+1}$ to a $\mathcal{C}_{n+1}$. }
  \label{Fig1}
\end{figure}

A cluster $\mathcal{C}_n$ consisting of $n$ bumps can be obtained by
using the following map on a unit circle
\begin{equation}\label{Eq6}
    \Phi_{n}(w) = \phi_{\lambda_{1},\theta_{1}} \circ
    \phi_{\lambda_{2},\theta_{2}} \circ \cdots \circ
    \phi_{\lambda_{n},\theta_{n}} (w),
\end{equation}
which corresponds to the following recursive relation for a cluster
$\mathcal{C}_{n+1}$ (see Fig. ¬\ref{Fig1}),
\begin{equation}\label{Eq7}
    \Phi_{n+1}(w) = \Phi_{n}(\phi_{\lambda_{n+1},\theta_{n+1}}(w)).
\end{equation}
Since $z=\Phi_{n}(w)$, one can obtain that
\begin{equation}\label{Eq8}
     dl = |\Phi^{'}_{n}(e^{i\theta})|d\theta,
\end{equation}
where the prime denotes for differentiation. \\In order to have
fixed-size bumps on the boundary of the cluster, since the linear
dimension at point $w$ is proportional to $|\Phi^{'}_{n}(w)|^{-1}$,
one obtains
\begin{equation}\label{Eq9}
    \lambda_{n+1} =
    \frac{\lambda_{0}}{|\Phi^{'}_{n}(e^{i\theta_{n+1}})|^2}.
\end{equation}
From Eq. ¬\ref{Eq2} and Eq. ¬\ref{Eq8} can be obtained that
\begin{equation}\label{Eq10}
    dP= |\nabla \psi||\Phi^{'}| d\theta = d \theta,
\end{equation}
indicating that the numbers $\theta_n$ have a uniform distribution
in the interval $0\leqslant\theta\leqslant2\pi$.

In this paper our analysis is based on the boundary of the clusters
and we need to have a uniform data on the boundary. This can be done
formally by using a uniform series of
$\left\{\beta_s\right\}_{s=1}^S$ during the conformal mapping from a
unit circle to the boundary of the cluster i.e.,
$\left\{w_s=e^{i\beta_s}\right\}_{s=1}^{S}$. This procedure can not
be applied operationally, because in order to have a reasonable data
in the fjords, one has to set $S\gg n$ which needs very long
simulation time.\\
Barra \emph{et al.}, \cite{Ref10} have focused on the branch points
of the map and introduced another approach for selecting the series
$\left\{\beta_s\right\}$. Following their approach, we define
$w_n^R$ and $w_n^L$ as "\emph{Right}" and "\emph{Left}" branch
points of the function $\phi_{\lambda_n, \theta_n}$ in the following
map, respectively
\begin{equation}\label{Eq11}
    e^{i\alpha_n^{R,L}}=\phi_{\lambda_n,\theta_n}(w_n^{R,L}),
\end{equation}
where $|\alpha_n^R-\alpha_n^L|/2\pi$ is the fraction of the unit
circle covered by the bump. Each new bump creates two new branch
points on the boundary and in case of probable overlapping with
previous branch point, some of the older ones will be removed. So
the maximum number of branch points will be $2n$. If $w_k^{R,L}$ be
a branch point of the $k$th bump without overlapping by the next
$(n-k)$ bumps, it would be an exposed branch point of the map
$\Phi_n$ but the pre-image of the branch on the unit circle will
change from $w_k^{R,L}$ to $w_{k,n}^{R,L}$
\begin{equation}\label{Eq12}
    \Phi_k(w_k^{R,L})=\Phi_n(w_{k,n}^{R,L}),
\end{equation}
such that
\begin{equation}\label{Eq13}
    w_{k,n}^{R,L}=\phi_{\lambda_n,\theta_n}^{-1}\circ\cdots\circ\phi_{\lambda_{k+1},\theta_{k+1}}^{-1}\left(w_k^{R,L}\right).
\end{equation}
The solvability of Eq. ¬\ref{Eq13} determines whether the branch
point remains exposed, and then by mapping them one gets a
reasonable image of the fjords.

\begin{figure}[b]
\includegraphics[width=0.45\textwidth]{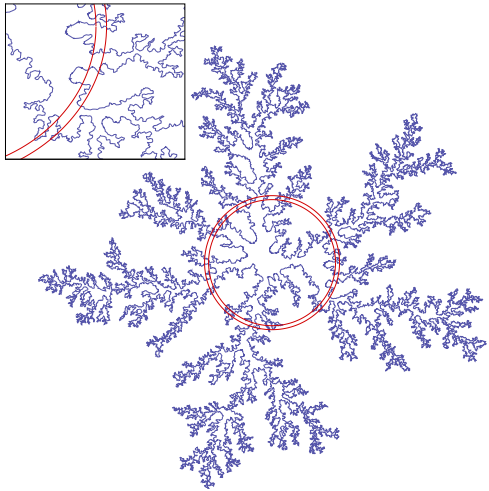}
\caption{Boundary of a typical simulated DLA cluster consisting of
$N=10^5$ bumps generated by using the HL algorithm, with
$a=\frac{1}{2}$. The plotted shell is used to study the muliscaling
properties of the boundary in Sec. ¬\ref{Boundary-Analysis}. The
width of the shell is magnified by a factor of $10$. Inset: a
close-up view of the cluster. } \label{Fig2}
\end{figure}

\section{Simulation}

The simulation of the boundary of DLA clusters of different sizes is
carried out using the algorithm discussed in the previous section.
We set the parameter $a=\frac{1}{2}$, for which the function
$\phi_{\lambda,\theta}(w)$ is analytically invertible.

At the $n$th step, $\theta_n$ and $\lambda_n$ are determined as
follows. $\theta_n$ is selected from a uniform distribution in the
range $[0,2\pi]$, and then $\lambda_n$ is computed using the Eq.
{¬\ref{Eq9}}. After determination of $\lambda$s and $\theta$s and
computing exposed branch points $w_{k,n}^{L,R}$, together with Eq.
¬\ref{Eq6}, the boundary of each cluster is determined.
\\We generated $2000$ clusters of number of bumps $10^3\leqslant
N\leqslant5\times10^4$ and $200$ clusters of $N=10^5$. A typical
growth cluster is shown in Fig. ¬\ref{Fig2}. All average quantities
which will be discussed later are taken over the simulated cluster
ensemble.

\section{Direct cluster analysis}

In this section we do some direct measurements based on the geometry
of clusters obtained from simulation. These include computation of
the fractal dimension of generated DLA clusters and size-dependence
of the variance of gyration radius of the clusters. We find a good
agreement between our results and ones obtained from the analytical
approach in \cite{Ref11,Ref11-}. We also measure the density
correlation function$-$which, to our knowledge, has not been
computed yet for the HL method$-$ and we investigate its dependence
on the size of the cluster. We find that the large-size behavior of
the function corresponds to an expected correlation exponent
$\alpha$ which is in a good agreement with the computed fractal
dimension.

\subsection{Scaling of Gyration Radius for DLA Cluster}

\begin{figure}[!htb]
\includegraphics[width=0.45\textwidth]{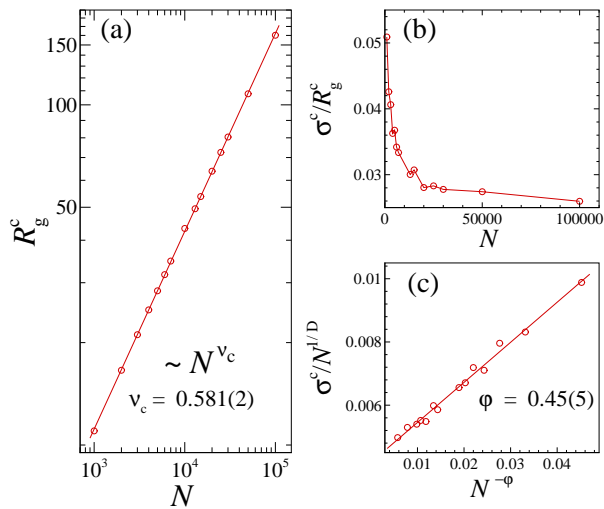}
\caption{(a) The average gyration radius of clusters $R_g^c$, versus
the number of bumps $N$. (b) Rescaled standard deviation of gyration
radius $\sigma^c/R^c_g$ versus $N$. (c) Rescaled standard deviation
of gyration radius ($\sigma^c/N^{1/D}$, $D=1.711$) versus
$N^{-\varphi}$. The error bars are almost in the same size as the
symbols.} \label{Fig3}
\end{figure}

The fractal dimension $D_c$ of DLA clusters generated by HL method
has been previously computed from the Laurent expansion of the
conformal map, cf. Eq. ¬\ref{Eq6}, equal to $D_c=1.713(3)$
\cite{Ref11, Ref11-}. The error in the last digit is indicated in
parentheses. This has been obtained from the scaling relation
between the first coefficient of the Laurent series of
$\phi_n(w)$ and the size of DLA cluster.\\
Since the first coefficient is proportional to the radius of the
cluster, this motivates us to measure the fractal dimension directly
using the scaling relation between the average gyration radius
$R^c_g$ of the cluster and the number of bumps $-$or equivalently
the cluster size$-$ $N$, i.e., $R^c_g\sim N^{\nu_c}$, where
$\nu_c=1/D_c$.\\The result is shown in Fig. \ref{Fig3}(a). We find
that $\nu_c=0.581(2)$, in good agreement with previous results.

Another important result pointed out in \cite{Ref12} is the
sharpness of the distribution of the first laurent coefficient. It
has been shown numerically that the rescaled distribution width of
squared first-laurent coefficient tends to zero as $N$ goes to
infinity. Here, we check the same idea for the gyration radius of
the clusters. The standard deviation of gyration radius is
calculated from $\sigma^c = \sqrt{\langle {R^c_g}^2\rangle-\langle
R^c_g\rangle^2}$, where $\langle\cdot\rangle$ denotes the ensemble
average over simulated clusters of size $N$. The rescaled $\sigma^c$
as a function of $N$ is plotted in Fig. \ref{Fig3}(b). As can be
seen from this figure the fluctuation tends to zero for larger
cluster size. This suggests that the rescaled distribution function
of gyration radius of the clusters tends asymptotically to
a $\delta$ function. \\
In order to investigate the asymptotic scaling behavior of
$\sigma^c$, we proceed in the same way as \cite{Ref13, Ref9}, where
the authors suggest that all of the length scales $\ell$ in DLA have
a scaling relation with $N$, like
\begin{equation}\label{Eq14}
\ell \sim N^{1/D}(a+bN^{-\varphi})
\end{equation} with
a single universal exponent $\varphi =0.33(6)$. \\Our computation
shown in Fig. \ref{Fig3}(c) agrees with this scaling relation but
with a different exponent of $\varphi =0.45(5)$, indicating that in
the limit $N\rightarrow \infty$, the fluctuation of gyration radius
has an asymptotic scaling behavior as that of the gyration radius,
and nevertheless, the exponent seems not to be universal (this will
be confirmed again in the following section for other length
scales).

\subsection{Density Correlation Function}

In this subsection, we compute the two-point correlation function
$c(\textbf{r})$, defined as
\begin{equation}\label{Eq15}
   c(\textbf{r})=\frac{1}{V}\sum_{\textbf{r}'}{\rho(\textbf{r}+\textbf{r}')\rho(\textbf{r}')},
\end{equation}
where $\rho(\textbf{r})$ is density at position $\textbf{r}$, and
the average is taken over all the points that belong to the cluster.
For isotropic clusters the density correlation depends only on
distance $r$.\\For self-similar fractals, $c(r)$ should have the
scaling form of $c(r)\sim r^{-\alpha}$, where the exponent $\alpha$
is named \emph{co-dimensionality} and is equal to $\alpha=d-D_c$,
where $d$ is the embedding dimension.

\begin{figure}[!htb]
\includegraphics[width=0.45\textwidth]{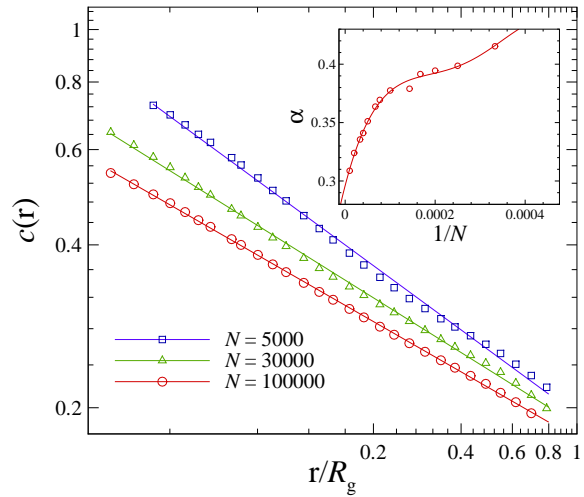}
\caption{Two-point density correlation function $c(r)$ for three
different cluster sizes $N$. The difference in the slope of the
solid lines indicates the size dependence of the correlation
exponent $\alpha$. The graph for $N=10^5$ is shifted downward by
$0.12$. Inset: the exponent $\alpha$ versus $1/N$. The solid line is
a polynomial fit of order $5$, which yields the asymptotic value of
$\alpha=0.29(1)$. The error bars are almost in the same size as the
symbols.} \label{Fig4}
\end{figure}

Operationally, we proceed as follows to determine the function
$c(r)$. For each sample in the ensemble of clusters of a fixed size,
we cover the cluster by a two dimensional square lattice. Then for
each lattice site belonging to the cluster, we consider an annulus
around it with mean radius of $r$ and thickness of a lattice
spacing. The density of the cluster points in the annulus is then
proportional to the two-point correlation function at distance $r$.
The average is then taken over both all lattice points in the
cluster and all clusters in the ensemble. This procedure is repeated
for annulus of different mean radius.\\
We find that, for intermediate distances, the function $c(r)$
exhibits a power-law behavior with an exponent $\alpha$ depending on
the cluster size $N$. This behavior is shown in Fig. ¬\ref{Fig4} for
three different sizes. The values of the exponent $\alpha$ as a
function of the inverse size of the cluster is depicted in the inset
of Fig. ¬\ref{Fig4}. In order to determine the value of the exponent
in the large-size limit, we fit a polynomial curve to the data. We
find that it extrapolates to $\alpha=0.29(1)$, whose value is
checked not to be affected by the degree of the fitted polynomial.
This value is in good agreement with the aforementioned relation
$\alpha=d-D_c$, with $D_c\sim 1.71$.

\section{Boundary Analysis \label{Boundary-Analysis}}

In this section, we study the scaling properties of various length
scales related to the boundary of DLA clusters produced by HL
method. We find that the fractal dimension of the boundary is the
same as the DLA cluster, in agreement with the same conclusion
reported in \cite{Ref14}, for DLA clusters produced according to the
original definition. We also check the simple scaling relation Eq.
\ref{Eq14} for various length scales including the gyration radius
$R_g^b$, maximum radius $R_{max}$ and width $R_w$ of the boundary
and their fluctuations. We find that all these length scales do not
obey the scaling form Eq. \ref{Eq14} with a single exponent
$\varphi$.\\Finally, we check the multiscaling hypothesis for the
boundary of the clusters and we will present evidence pointing to
the existence of such anomalous scaling.

\subsection{Scaling of boundary characteristic lengths}

Each cluster boundary is divided into segments such that $i$th
segment has a length $l_i$, and the distance of the midpoint of the
segment from the center of mass is denoted by $R_i$. During the
calculations, this procedure attributes a weight of $l_i$ to each
distance $R_i$ and measures the following length scales in a more
delicate manner.

\subsubsection{Gyration radius of the boundary, $R_g^b$}

\begin{figure}[!t]
\includegraphics[width=0.45\textwidth]{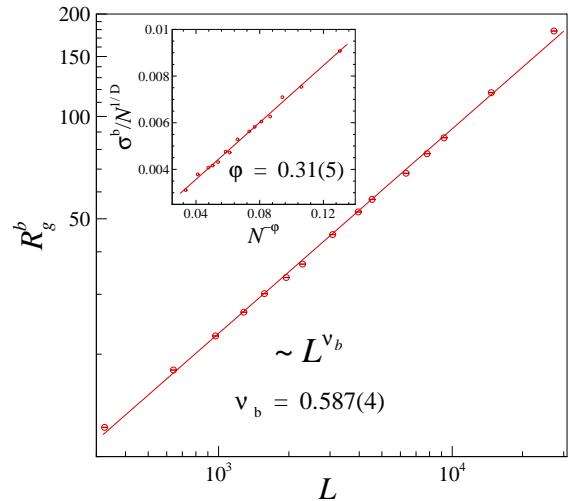}
\caption{Average gyration radius of the boundary $R^b_g$, versus the
average length of the boundary $L$. Inset: rescaled standard
deviation of the gyration radius ($\sigma^b/N^{1/D}$, $D=1.711$)
versus $N^{-\varphi}$. The error bars are almost in the same size as
the symbols.} \label{Fig5}
\end{figure}

The gyration radius of the boundary $R_g^b$ is defined by
$R_g^b=\sqrt{\frac{1}{L}\sum_i {l_i R_i^2}}$, where $L$ is the total
length of the boundary, and the sum runs over all segments on it.
The fractal dimension of the boundary $D_b$ can be measured by using
the scaling relation $R_g^b\sim L^{\nu_b}$, where $\nu_b=1/D_b$.
Fig. ¬\ref{Fig5} shows the ensemble average of gyration radius
versus the average length of the boundary. We find that
$\nu_b=0.587(4)$. This indicates that within the statistical errors,
a DLA cluster generated by HL method and its boundary have a same
fractal dimension i.e., $\nu_c=\nu_b$. This result is the same as
one obtained before for DLA patterns grown according to the original
definition \cite{Ref14}. It may be considered as another evidence
that the patterns generated by method of iterated conformal maps
proposed by Hastings and Levitov agree statistically with ones
originally introduced by Witten and Sanders. We have also checked
the scaling of $R_g^b$ with the cluster size and we found the same
behavior as $R_g^c$ with $N$.\\The inset of Fig. ¬\ref{Fig5} shows
the plot of the rescaled standard deviation of $R_g^b$ i.e.,
$\sigma^b/N^{1/D_c}$ against $N^{-\varphi}$. We find that
$\varphi=0.31(5)$, in agreement with Eq. \ref{Eq14}.

\subsubsection{Maximum radius of the boundary, $R_{max}$}

\begin{figure}[!b]
\includegraphics[width=0.45\textwidth]{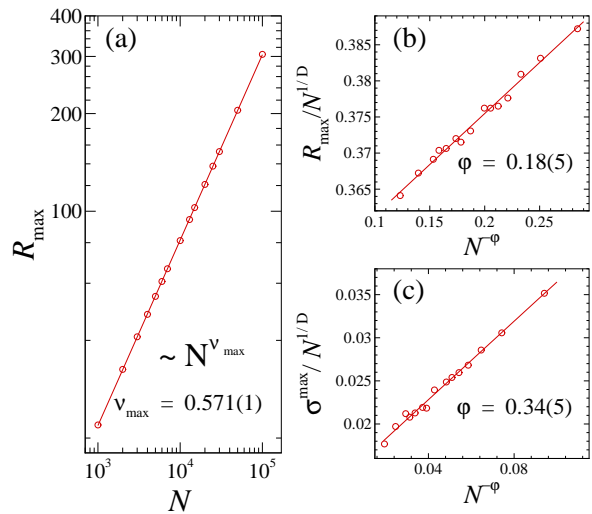}
\caption{(a) The ensemble average of the most-far boundary segment
from the seed $R_{max}$, versus the number of bumps $N$. (b)
Rescaled $R_{max}$ (i.e., $R_{max}/N^{1/D}$, $D=1.711$), versus
$N^{-\varphi}$. (c) Rescaled standard deviation of $R_{max}$ (i.e.,
$\sigma^{max}/N^{1/D}$, $D=1.711$) versus $N^{-\varphi}$. The error
bars are almost in the same size as the symbols.} \label{Fig6}
\end{figure}

The other length scales we discuss here, are the lengths related to
the maximum value of $R_i$ in each cluster boundary represented by
$R_{max}$ in Fig. ¬\ref{Fig6}. We observe from Fig. ¬\ref{Fig6}(a)
that the ensemble average of $R_{max}$ scales with size $N$, with
$\nu_{max} = 0.571(1)$, different from the gyration radius exponent.
As shown in Fig. ¬\ref{Fig6}(b), the rescaled $R_{max}$ follows the
simple scaling behavior of Eq. \ref{Eq14}, with a quite different
exponent of $\varphi=0.18(5)$ from the proposed \emph{universal}
value of $\varphi=0.33(6)$ in \cite{Ref13, Ref9}. We also checked
this simple scaling behavior for the rescaled standard deviation of
$R_{max}$, in agreement with Eq. \ref{Eq14} (see Fig.
¬\ref{Fig6}(c)).

\subsubsection{Interface width, $R_w$}

\begin{figure}[!h]
\includegraphics[width=0.45\textwidth]{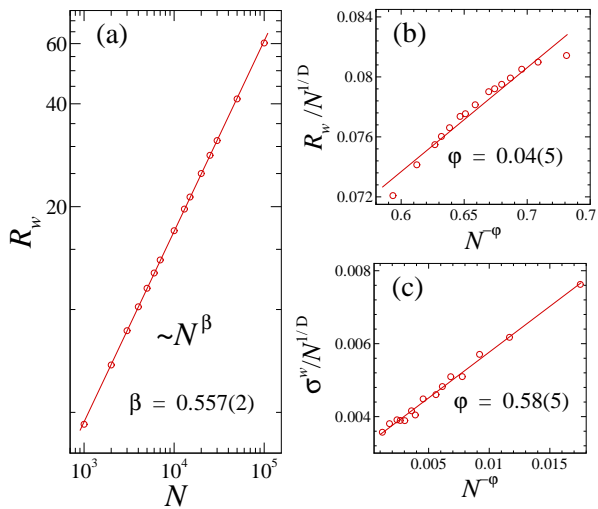}
\caption{(a) The average interface width of the boundary of cluster
$R_w$, versus the number of bumps $N$. (b) Rescaled interface width
(i.e., $R_w/N^{1/D}$, $D=1.711$) versus $N^{-\varphi}$. (c) Rescaled
standard deviation of the interface width (i.e., $\sigma^w/N^{1/D}$,
$D=1.711$) versus $N^{-\varphi}$. The error bars are almost in the
same size as the symbols.} \label{Fig7}
\end{figure}

According to the analogy between the DLA growing cluster and
non-Euclidean growing interfaces, the interface width $R_w$ can be
defined by $R_w= \sqrt{\frac{1}{L}\sum _i {l_i(R_i -\bar{R})^2}}$,
where the mean radius of the cluster is $\bar{R}=\frac{1}{L}\sum _i
{l_iR_i}$. \\ The growth exponent $\beta$ can be obtained from the
evolution of the interface width $R_w\sim N^\beta$. As shown in Fig.
\ref{Fig7}(a), we obtain the growth exponent for DLA clusters
generated by HL method equal to $\beta=0.557(2)$. We checked the
correction to scaling for the exponent, according to Eq. \ref{Eq14}
represented in Fig. \ref{Fig7}(b), and we conclude that no
correction exists. The fluctuation of the interface width (see Fig.
\ref{Fig7}(c)) exhibits a simple scaling relation of form Eq.
\ref{Eq14}, with a correction to scaling exponent of
$\varphi=0.58(5)$. This exponent is very different from those
obtained for hitherto mentioned length scales, and far from its
proposed universal value.

The scaling properties of the interface width, apparently deviates
from the simple scaling of Eq. \ref{Eq14}, which has been proposed
in \cite{Ref9} on refuting the multiscaling property of DLA cluster.
The deviations of these boundary related length scales from the
simple scaling behavior, motivated us to check an extension of the
multiscaling property (previously applied for the mass of DLA
clusters) to the length of the perimeter of clusters.


\subsection{Multiscaling analysis of the boundary of DLA clusters}

In this section, we extend the concept of multiscaling, previously
used for the mass of the DLA clusters \cite{ Ref15, Ref16, Ref17,
Ref18}, to the length of the border of DLA. Our measurement for the
perimeter of DLA clusters of size up to $10^5$ particles (or bumps),
reveals the multiscaling behavior of the border. \\ For each cluster
size, we generated an ensemble of DLA clusters by using the HL
method and the perimeter of each sample has been determined as
described in Sec. \ref{section HL}. We proceed as follows: for each
sample perimeter in the ensemble of size $N$ and average gyration
radius of $R_g^b$, a shell of radius $r$ and of width $dr$ (which is
about the linear size of a bump) is drawn (see Fig. \ref{Fig2} for
illustration). Then we measure the density profile
$g\left(r,R_g\right)$ defined as

\begin{equation}\label{Eq16}
        g\left(r,R_g\right)dr = dl,
\end{equation}
where $dl$ is the total length of the boundary within the shell of
radius $r$.

\begin{figure}[!t]
\includegraphics[width=0.42\textwidth]{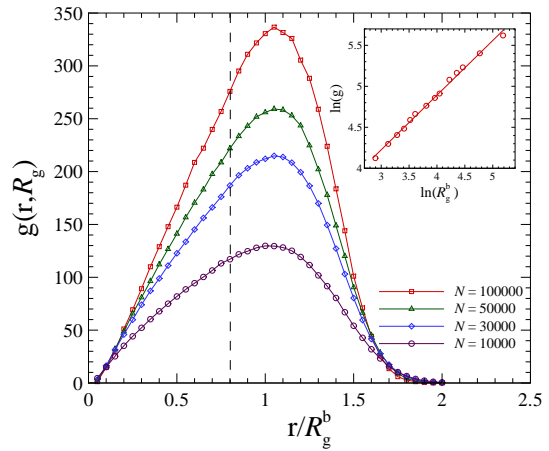}
\caption{Examples of the density profile of the boundary length of
the clusters within a shell of rescaled radius $x=r/R_g^b$,
represented for four different sizes. Inset: log-log plot of the
density profile at a certain rescaled radius of $x=r/R_g^b=0.8$ (the
dashed line in the main figure). The slope of the fitted solid line
yields $D(x=0.8)=1.67$ for $N = 100000$. This figure summarizes the
procedure we applied to obtain the functions $D(x)$ in Fig.
\ref{Fig9}. The error bars are almost in the same size as the
symbols. } \label{Fig8}
\end{figure}

The plot of $g\left(r,R_g\right)$ as a function of the rescaled
radius $x=r/R_g^b$, within $0.1\leq x\leq 2$, is shown in Fig.
¬\ref{Fig8} for four different sizes. This function has a maximum
for distances around the gyration radius of the cluster. Assuming
the scale invariance of the density profile \cite{Ref16}, the
multiscaling exponent $D(x)$ can be defined as
\begin{equation}\label{Eq17}
        g\left(r,R_g\right) = C(x) R_g^{D(x)-1},
\end{equation}
where $C(x)$ is a scaling function. Thus, the multiscaling exponent
can be obtained using the following relation
\begin{equation}\label{Eq18}
        D(x) = 1 + \left.\frac{\partial \ln g\left(r,R_g\right)}{\partial \ln
        R_g}\right|_x \,.
\end{equation}

The inset of Fig. ¬\ref{Fig8} shows the procedure we used to
determine the multiscaling exponent as a function of $x$. At each
$x$, the values of the density profile are read from Fig.
¬\ref{Fig8} for each cluster size of gyration radius $R_g^b$, and
then $D(x)$ is determined by Eq. \ref{Eq18}.

\begin{figure}[!h]
\includegraphics[width=0.42\textwidth]{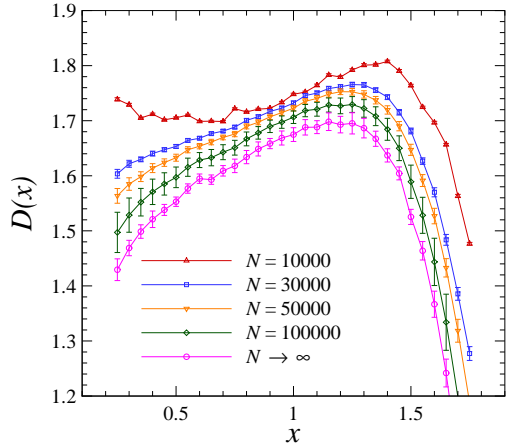}
\caption{Multiscaling fractal dimension D(x) of the boundary for
different cluster sizes as a function of $x = r/R_g^b$.}
\label{Fig9}
\end{figure}

The whole behavior of $D(x)$ for different size intervals is shown
in Fig. ¬\ref{Fig9}. This shows that the function $D(x)$ does not
tend to a constant value as size increases, and there is a maximum
around $x\simeq1.2$ whose location does not depend on the size of
cluster. Using the curves of Fig. ¬\ref{Fig9} (and other similar
curves obtained for other cluster sizes which not shown in the
figure), we also estimated the value of $D(x)$ at each $x$ in the
limit of $N\rightarrow \infty$. As shown in Fig. ¬\ref{Fig9}, $D(x)$
is not constant and varies with $x$, suggesting a multiscaling
behavior. We therefore conclude that the perimeter of the DLA
clusters generated by HL method does not have simple scaling, and
thus a set of scaling exponents is needed to be described.

\section{Conclusion}

We studied scaling properties of DLA clusters generated by the
Hastings-Levitov method. First, we calculated the fractal dimension
of the clusters by direct analyzing of the DLA patterns in agreement
with the previous results. We also computed the two-point
correlation function of the mass of the cluster, and we found that
in the large-size limit, it agrees with the obtained fractal
dimension.

In the second part of the paper, we focused on the border of the DLA
clusters and we investigated their scaling properties. We found that
the fractal dimension of the perimeter is equal to that of the
cluster. We checked the simple scaling behavior for various length
scales including the gyration radius, maximum radius and the
interface width of the boundary, together with their fluctuations.
We found that all of these length scales do not have a simple
scaling with a universal correction to scaling exponent. The growth
exponent has been obtained from the evolution of the interface
width. Finally, we found that the perimeter of DLA displays an
asymptotic multiscaling property.

\pagebreak


\begin{thebibliography}{00}


\expandafter\ifx\csname
natexlab\endcsname\relax\def\natexlab#1{#1}\fi
\expandafter\ifx\csname bibnamefont\endcsname\relax
  \def\bibnamefont#1{#1}\fi
\expandafter\ifx\csname bibfnamefont\endcsname\relax
  \def\bibfnamefont#1{#1}\fi
\expandafter\ifx\csname citenamefont\endcsname\relax
  \def\citenamefont#1{#1}\fi
\expandafter\ifx\csname url\endcsname\relax
  \def\url#1{\texttt{#1}}\fi
\expandafter\ifx\csname urlprefix\endcsname\relax\def\urlprefix{URL
}\fi \providecommand{\bibinfo}[2]{#2}
\providecommand{\eprint}[2][]{\url{#2}}







\bibitem[{\citenamefont{Witten, T.A. et~al.}(1981)}]{Ref1}
  \bibinfo{author}{\bibfnamefont{T.A.}~\bibnamefont{Witten}} and
  \bibinfo{author}{\bibfnamefont{L.M.}~\bibnamefont{Sander}},
  \bibinfo{Jurnal}{{Phys. Rev. Lett.}}
  \textbf{\bibinfo{volume}{47}}
  \bibinfo{pages}{1400}.
 (\bibinfo{year}{1981})



\bibitem[{\citenamefont{Niemeyer et~al.}(1984)}]{Ref2}
  \bibinfo{author}{\bibfnamefont{L.}~\bibnamefont{Niemeyer}},
  \bibinfo{author}{\bibfnamefont{L.}~\bibnamefont{Pietronero}},
  \bibinfo{author}{\bibfnamefont{H.J.}~\bibnamefont{Wiesmann}},
  \bibinfo{Jurnal}{{Phys. Rev. Lett.}}
  \textbf{\bibinfo{volume}{52}}
  \bibinfo{pages}{1033}.
 (\bibinfo{year}{1984})

\bibitem[{\citenamefont{Brady et~al.}(1984)}]{Ref3}
  \bibinfo{author}{\bibfnamefont{R.M.}~\bibnamefont{Brady}} and
  \bibinfo{author}{\bibfnamefont{R.C.}~\bibnamefont{Ball}},
  \bibinfo{Jurnal}{{Nature (London)}}
  \textbf{\bibinfo{volume}{309}}
  \bibinfo{pages}{225}.
 (\bibinfo{year}{1984})


\bibitem[{\citenamefont{Matsushita et~al.}(1984)}]{Ref4}
  \bibinfo{author}{\bibfnamefont{M.}~\bibnamefont{Matsushita}},
  \bibinfo{author}{\bibfnamefont{M.}~\bibnamefont{Sano}},
  \bibinfo{author}{\bibfnamefont{Y.}~\bibnamefont{Hayakawa}},
  \bibinfo{author}{\bibfnamefont{H.}~\bibnamefont{Honjo}},
  \bibinfo{author}{\bibfnamefont{Y.}~\bibnamefont{Sawada}},
  \bibinfo{Jurnal}{{Phys. Rev. Lett.}}
  \textbf{\bibinfo{volume}{53}}
  \bibinfo{pages}{286}.
 (\bibinfo{year}{1984})


\bibitem[{\citenamefont{Paterson et~al.}(1984)}]{Ref5}
  \bibinfo{author}{\bibfnamefont{L.}~\bibnamefont{Paterson}}
  \bibinfo{Jurnal}{{Phys. Rev. Lett.}}
  \textbf{\bibinfo{volume}{52}}
  \bibinfo{pages}{1621}.
 (\bibinfo{year}{1984})



\bibitem[{\citenamefont{}()}]{Ref6}
  \bibinfo{author}{\bibfnamefont{M.B.}~\bibnamefont{Hastings}} and
  \bibinfo{author}{\bibfnamefont{L.S.}~\bibnamefont{Levitov}},
  \bibinfo{Jurnal}{{Physica D}}
  \textbf{\bibinfo{volume}{47}}
  \bibinfo{pages}{244}.
 (\bibinfo{year}{1998})



\bibitem[{\citenamefont{}()}]{Ref7}
  \bibinfo{author}{\bibfnamefont{A.}~\bibnamefont{Coniglio}} and
  \bibinfo{author}{\bibfnamefont{M.}~\bibnamefont{Zannetti}},
  \bibinfo{Jurnal}{{Physica A}}
  \textbf{\bibinfo{volume}{163}}
  \bibinfo{pages}{325}.
 (\bibinfo{year}{1990})


\bibitem[{\citenamefont{}()}]{Ref13}
  \bibinfo{author}{\bibfnamefont{E.}~\bibnamefont{Somfai}},
  \bibinfo{author}{\bibfnamefont{L.M.}~\bibnamefont{Sander}},
  \bibinfo{author}{\bibfnamefont{R.C.}~\bibnamefont{Ball}},
  \bibinfo{Jurnal}{{Phys. Rev. Lett.}}
  \textbf{\bibinfo{volume}{83}}
  \bibinfo{pages}{5523-5526}.
 (\bibinfo{year}{1999})




\bibitem[{\citenamefont{}()}]{Ref8}
  \bibinfo{author}{\bibfnamefont{R.C.}~\bibnamefont{Ball}},
  \bibinfo{author}{\bibfnamefont{N.E.}~\bibnamefont{Bowler}},
  \bibinfo{author}{\bibfnamefont{L.M.}~\bibnamefont{Sander}},
  \bibinfo{author}{\bibfnamefont{E.}~\bibnamefont{Somfai}},
  \bibinfo{Jurnal}{{Phys. Rev. E}}
  \textbf{\bibinfo{volume}{66}}
  \bibinfo{pages}{026109}.
 (\bibinfo{year}{2002})


\bibitem[{\citenamefont{}()}]{Ref9}
  \bibinfo{author}{\bibfnamefont{E.}~\bibnamefont{Somfai}},
  \bibinfo{author}{\bibfnamefont{R.C.}~\bibnamefont{Ball}},
  \bibinfo{author}{\bibfnamefont{N.E.}~\bibnamefont{Bowler}},
  \bibinfo{author}{\bibfnamefont{L.M.}~\bibnamefont{Sander}},
  \bibinfo{Jurnal}{{Physica A}}
  \textbf{\bibinfo{volume}{325}}
  \bibinfo{pages}{19}.
 (\bibinfo{year}{2003})




\bibitem[{\citenamefont{}()}]{Ref10}
  \bibinfo{author}{\bibfnamefont{F.}~\bibnamefont{Barra}},
  \bibinfo{author}{\bibfnamefont{B.}~\bibnamefont{Davidovitch}},
  \bibinfo{author}{\bibfnamefont{I.}~\bibnamefont{Procaccia}},
  \bibinfo{Jurnal}{{Phys. Rev. E}}
  \textbf{\bibinfo{volume}{65}}
  \bibinfo{pages}{046144}.
 (\bibinfo{year}{2002})




\bibitem[{\citenamefont{}()}]{Ref11}
  \bibinfo{author}{\bibfnamefont{B.}~\bibnamefont{Davidovitch}} and
  \bibinfo{author}{\bibfnamefont{I.}~\bibnamefont{Procaccia}},
  \bibinfo{Jurnal}{{Phys. Rev. Lett.}}
  \textbf{\bibinfo{volume}{85}}
  \bibinfo{pages}{3608}.
 (\bibinfo{year}{2000})


\bibitem[{\citenamefont{}()}]{Ref11-}
  \bibinfo{author}{\bibfnamefont{B.}~\bibnamefont{Davidovitch}},
    \bibinfo{author}{\bibfnamefont{A.}~\bibnamefont{Levermann}},
  \bibinfo{author}{\bibfnamefont{I.}~\bibnamefont{Procaccia}},
  \bibinfo{Jurnal}{{Phys. Rev. E}}
  \textbf{\bibinfo{volume}{62}}
  \bibinfo{pages}{R5919}.
 (\bibinfo{year}{2000})

\bibitem[{\citenamefont{}()}]{Ref12}
  \bibinfo{author}{\bibfnamefont{B.}~\bibnamefont{Davidovitch}},
  \bibinfo{author}{\bibfnamefont{H.G.E.}~\bibnamefont{Hentchel}},
  \bibinfo{author}{\bibfnamefont{Z.}~\bibnamefont{Olami}},
  \bibinfo{author}{\bibfnamefont{}~\bibnamefont{et al.}},
  \bibinfo{Jurnal}{{Phys. Rev. E}}
  \textbf{\bibinfo{volume}{59}}
  \bibinfo{pages}{1368-1378}.
 (\bibinfo{year}{1999})





 \bibitem[{\citenamefont{}()}]{Ref14}
  \bibinfo{author}{\bibfnamefont{C.}~\bibnamefont{Amitrano}},
  \bibinfo{author}{\bibfnamefont{P.}~\bibnamefont{Meakin}},
  \bibinfo{author}{\bibfnamefont{H.E.}~\bibnamefont{Stanley}},
  \bibinfo{Jurnal}{{Phys. Rev. A}}
  \textbf{\bibinfo{volume}{40}}
  \bibinfo{pages}{1713}.
 (\bibinfo{year}{1989})



\bibitem[{\citenamefont{}()}]{Ref15}
  \bibinfo{author}{\bibfnamefont{M.}~\bibnamefont{Plischke}} and
  \bibinfo{author}{\bibfnamefont{Z.}~\bibnamefont{Racz}},
  \bibinfo{Jurnal}{{Phys. Rev. Lett.}}
  \textbf{\bibinfo{volume}{53}}
  \bibinfo{pages}{415-418}.
 (\bibinfo{year}{1984})

 \bibitem[{\citenamefont{}()}]{Ref16}
  \bibinfo{author}{\bibfnamefont{C.}~\bibnamefont{Amitrano}},
  \bibinfo{author}{\bibfnamefont{A.}~\bibnamefont{Coniglio}},
  \bibinfo{author}{\bibfnamefont{P.}~\bibnamefont{Meakin}},
  \bibinfo{author}{\bibfnamefont{M.}~\bibnamefont{Zannetti}},
  \bibinfo{Jurnal}{{Phys. Rev. B}}
  \textbf{\bibinfo{volume}{44}}
  \bibinfo{pages}{4974}.
 (\bibinfo{year}{1991})


\bibitem[{\citenamefont{}()}]{Ref17}
  \bibinfo{author}{\bibfnamefont{B.B.}~\bibnamefont{Mandelbrot}} and
  \bibinfo{author}{\bibfnamefont{B.}~\bibnamefont{Kol}},
  \bibinfo{Jurnal}{{Phys. Rev. Lett.}}
  \textbf{\bibinfo{volume}{88}}
  \bibinfo{pages}{055501}.
 (\bibinfo{year}{2002})





\bibitem[{\citenamefont{}(1998)}]{Ref18}
  \bibinfo{author}{\bibfnamefont{A.Y.}~\bibnamefont{Menshutin}} and
  \bibinfo{author}{\bibfnamefont{L.N.}~\bibnamefont{Shchur}},
  \bibinfo{Jurnal}{{Phys. Rev. E}}
  \textbf{\bibinfo{volume}{73}}
  \bibinfo{pages}{011407}.
 (\bibinfo{year}{2006})




\end{thebibliography}
\end{document}